\documentclass[journal,twocolumn]{IEEEtran}
\usepackage[noadjust]{cite}

\usepackage[T1]{fontenc}
\usepackage{mathpazo}

\usepackage{graphicx}
\graphicspath{{figures/}{./}} 
\makeatletter
\def\maxwidth{\ifdim\Gin@nat@width>\linewidth\linewidth
\else\Gin@nat@width\fi}
\makeatother
\let\Oldincludegraphics\includegraphics
\renewcommand{\includegraphics}[1]{\Oldincludegraphics[width=.8\maxwidth]{#1}}
\usepackage{caption}
\DeclareCaptionLabelFormat{nolabel}{}
\usepackage{adjustbox} 
\usepackage{xcolor} 
\usepackage{enumerate} 
\usepackage{geometry} 
\usepackage{amsmath} 
\usepackage{amssymb} 
\usepackage{textcomp} 
\AtBeginDocument{%
}
\usepackage{upquote} 
\usepackage{eurosym} 
\usepackage[mathletters]{ucs} 
\usepackage[utf8x]{inputenc} 
\usepackage{fancyvrb} 
\usepackage{grffile} 
\usepackage{hyperref}
\usepackage{longtable} 
\usepackage{booktabs}  
\usepackage[inline]{enumitem} 
\usepackage[normalem]{ulem} 
\usepackage{mathrsfs}
    
\usepackage{mathptmx}
\usepackage{algorithm}\usepackage{algorithmic}
\usepackage{braket}

\usepackage{subcaption}

\definecolor{urlcolor}{rgb}{0,.145,.698}
\definecolor{linkcolor}{rgb}{.71,0.21,0.01}
\definecolor{citecolor}{rgb}{.12,.54,.11}

\definecolor{ansi-black}{HTML}{3E424D}
\definecolor{ansi-black-intense}{HTML}{282C36}
\definecolor{ansi-red}{HTML}{E75C58}
\definecolor{ansi-red-intense}{HTML}{B22B31}
\definecolor{ansi-green}{HTML}{00A250}
\definecolor{ansi-green-intense}{HTML}{007427}
\definecolor{ansi-yellow}{HTML}{DDB62B}
\definecolor{ansi-yellow-intense}{HTML}{B27D12}
\definecolor{ansi-blue}{HTML}{208FFB}
\definecolor{ansi-blue-intense}{HTML}{0065CA}
\definecolor{ansi-magenta}{HTML}{D160C4}
\definecolor{ansi-magenta-intense}{HTML}{A03196}
\definecolor{ansi-cyan}{HTML}{60C6C8}
\definecolor{ansi-cyan-intense}{HTML}{258F8F}
\definecolor{ansi-white}{HTML}{C5C1B4}
\definecolor{ansi-white-intense}{HTML}{A1A6B2}
\definecolor{ansi-default-inverse-fg}{HTML}{FFFFFF}
\definecolor{ansi-default-inverse-bg}{HTML}{000000}

\definecolor{incolor}{rgb}{0.0, 0.0, 0.5}
\definecolor{outcolor}{rgb}{0.545, 0.0, 0.0}

\DefineVerbatimEnvironment{Highlighting}{Verbatim}{commandchars=\\\{\}}

\let\Oldtex\TeX
\let\Oldlatex\LaTeX
\renewcommand{\TeX}{\textrm{\Oldtex}}
\renewcommand{\LaTeX}{\textrm{\Oldlatex}}

\def\eqref#1{Eq.~(\ref{#1})}
\def\figref#1{Figure~\ref{#1}}
\def\secref#1{Section~\ref{#1}}

\def\gb{\gamma,\beta}

\title{Evaluation of QAOA based on the approximation ratio of individual samples}

\input{pygments-definitions}

\hypersetup{
  breaklinks=true,  
  colorlinks=true,
  urlcolor=urlcolor,
  linkcolor=linkcolor,
  citecolor=citecolor,
  }

\begin{document}

\renewcommand{\figurename}{Figure}
\captionsetup{labelformat=simple}

\author{\IEEEauthorblockN{Jason Larkin\IEEEauthorrefmark{1}, Matías Jonsson\IEEEauthorrefmark{1}\IEEEauthorrefmark{2}, Daniel Justice\IEEEauthorrefmark{1}, and Gian Giacomo Guerreschi\IEEEauthorrefmark{3}}\\ \vspace{0.05in}
\IEEEauthorblockA{\IEEEauthorrefmark{1}Software Engineering Institute, Carnegie Mellon University, Pittsburgh, Pennsylvania 15213}\\\IEEEauthorblockA{\IEEEauthorrefmark{2}Department of Physics, Carnegie Mellon University, Pittsburgh, Pennsylvania, 15213}\\\IEEEauthorblockA{\IEEEauthorrefmark{3}Intel Labs, Santa Clara, California 95054}}

\maketitle

\begin{abstract}

The Quantum Approximate Optimization Algorithm (QAOA) is a hybrid quantum-classical algorithm to solve binary-variable optimization problems. Due to the short circuit depth and its expected robustness to systematic errors, it is one of the promising candidates likely to run on near-term quantum devices.
We simulate the performance of QAOA applied to the Max-Cut problem and compare it with some of the best classical alternatives, for exact, approximate and heuristic solution. When comparing solvers, their performance is characterized by the computational time taken to achieve a given quality of solution. Since QAOA is based on sampling, we utilize performance metrics based on the probability of observing a sample above a certain quality. In addition, we show that the QAOA performance varies significantly with the graph type.
By selecting a suitable optimizer for the variational parameters and reducing the number of function evaluations, QAOA performance improves by up to 2 orders of magnitude compared to previous estimates. Especially for 3-regular random graphs, this setting decreases the performance gap with classical alternatives. Because of the evolving QAOA computational complexity-theoretic guidance, we suggest a framework for the search for quantum advantage which incorporates a large number of problem instances and all three classical solver modalities: exact, approximate, and heuristic. 
\end{abstract}

\section{Introduction}
\label{sec:introduction}

The challenge for near-term noisy intermediate scale quantum computing (NISQ) is to demonstrate quantum advantage, where some computation is performed by a quantum computer that is  classically computationally intractable \cite{preskill_quantum_2018}. While ``quantum supremacy'' may have been demonstrated for the task of sampling the outcome of random quantum circuits \cite{arute_quantum_2019,boixo_characterizing_2018,neill_blueprint_2018,farhi_quantum_2019-1,bremner_achieving_2017}, practical applications of post-classical computation still need to be demonstrated.

A common practical problem used as a benchmark for both high performance classical and quantum computing is Max-Cut, a graph partition problem with applications in domains such as machine scheduling \cite{alidaee_0-1_1994}, image recognition \cite{neven_image_2008}, electronic circuit layout \cite{deza_applications_1994}, and software verification$/$validation \cite{jose_cause_2011, guo_complexity_2013}. Max-Cut is a NP-hard problem\cite{garey_computers_2009} and is naturally phrased as a Quadratic Unconstrained Binary Optimization (QUBO) problem.

Here, we investigate the Quantum Approximate Optimization Algorithm (QAOA) \cite{farhi_quantum_2014} applied to Max-Cut for different types of graphs. QAOA is a hybrid quantum-classical algorithm based on a circuit whose variational parameters are optimized with respect to the specific instance to solve. While QAOA is labelled an approximate solver, performance guarantees are known only for a small subset of problems and instance types \cite{wang_quantum_2018,hastings_classical_2019,bravyi_classical_2019,farhi_quantum_2020,farhi_quantum_2019}. In practice its performance is judged heuristically with the goal of approximating the solution of hard problems. We numerically simulate the execution of QAOA experiments and estimate the time required by noiseless hardware to provide candidate solutions above a certain approximation ratio. For instances of small size and reasonable convergence criteria, QAOA often returns the global solution.

We compare QAOA against a list of some of the best performing classical exact \cite{kugel_improved_nodate}, approximate \cite{goemans_improved_1995}, and heuristic  solvers \cite{dunning_what_2018} in terms of time-to-solution \cite{mandra_adiabatic_2015,mandra_strengths_2016} and quality for a range of graph types \cite{dunning_what_2018,farhi_quantum_2019,montanari_optimization_2019,moussa_quantum_2020}. One of the main contributions of this work is the definition of the framework for such comparison. From its original formulation, QAOA is expected to return the best candidate solution found in the course of the optimization. However, the objective that guides such optimization is an expectation value averaged over multiple candidate solutions. Most studies relied on concentration arguments to report the average value instead of the best single outcome \cite{farhi_quantum_2014,shaydulin_community_2018,shaydulin_evaluating_2019,crooks_performance_2018,otterbach_unsupervised_2017,brandao_for_2018}. In this work, given a desired approximation ratio, we formulate the performance of QAOA as the time needed before at least one sample with approximation ratio above the desired threshold is observed with probability at least 50\%. Alternative metrics have been proposed, for example see Refs \cite{barkoutsos_improving_2019, li_quantum_2020, kim_leveraging_2019}.

This paper is organized as follows: Section \ref{sec:qaoa} describes QAOA, its applications and some previous results. Section \ref{sec:methods} introduces the framework to report QAOA performance based on the approximation ratio of single samples. It also provides information about how we perform the numerical simulations and the classes of graphs we consider.  The results from the numerical simulations are compared with classical solvers in Section \ref{sec:results}, and the results are discussed in Sections \ref{sec:Discussion} and \ref{sec:conclusions}. Appendix \ref{sec:appendix} provides further data comparing classical solvers alone.

\section{Quantum Approximate Optimization Algorithm}
\label{sec:qaoa}

The Quantum Approximate Optimization Algorithm (QAOA) is a hybrid quantum-classical algorithm that variationally improves shallow quantum circuits to prepare states with desired properties. In its original formulation, the target state corresponds to the solution of binary-variable minimization problems \cite{farhi_quantum_2014}. The cost function is represented by a linear combination of operators that are diagonal in the Z basis:
%
\begin{equation}
\label{eq:cost-function}
    \hat{C} = \sum_{m=1}^{M} \hat{C}_m \; ,
\end{equation}
where $M$ is the number of terms, each proportional to the product of a few Z Pauli matrices. The observable $\hat{C}$ is obtained from the classical cost function by substituting classical binary variables $z_k\in\{-1,+1\}$ with quantum operators $\hat{Z}_k$, the Z Pauli matrix on qubit $k$. For Quadratic Unconstrained Binary Optimization (QUBO) problems, each term $\hat{C_m}$ is proportional to $\hat{Z}_i \hat{Z}_j$ for a certain pair $(i, j)$ of qubits.

The QAOA circuit is obtained by alternating two blocks of quantum operations for a total of $p$ times each, with $p$ being an adjustable parameter. Each block is characterized by a scalar parameter (denoted either by $\gamma_k$ or $\beta_k$) and their values uniquely determine the output state. The form of QAOA circuits is connected to the digital implementation of adiabatic quantum optimization \cite{mbeng_quantum_2019}.

We denote the variational state prepared by the QAOA circuit as:
\begin{equation}
\label{eq:variational-state}
    \ket{\gb} = e^{-i \beta_p \hat{B}} e^{-i \gamma_p \hat{C}} \dots\,
                         e^{-i \beta_1 \hat{B}} e^{-i\gamma_1 \hat{C}} \ket{+}^{\otimes N} \; ,
\end{equation}
where $B=\sum_k \hat{X}_k$ is the sum of the X Pauli matrices associated to each qubit, $\ket{+}=(\ket{0}+\ket{1})/\sqrt 2$, and $N$ is the number of qubits. The circuit's depth is linear in $p$ and the number of its parameters is $2p$. In this work, $(\gb)=(\gamma_p, \beta_p, \dots, \gamma_2, \beta_2, \gamma_1, \beta_1)$

QAOA has already been applied to a growing number of combinatorial optimization problems like community detection \cite{shaydulin_community_2018}, vertex cover \cite{cook_quantum_2019}, maximum  independent set \cite{farhi_quantum_2020}, and tail assignment \cite{vikstal_applying_2019}. However its most studied application is probably to a graph partition problem called Max-Cut. Its formulation is easily stated: divide the vertices of a graph in two groups and count the number of edges connecting vertices of different groups. These edges are usually said to be ``cut'' by the partition. The answer to the problem is returning the partition with the largest number of cut edges. Despite its simplicity, Max-Cut appears in many domains and is a NP-complete problem \cite{alidaee_0-1_1994,neven_image_2008,deza_applications_1994,jose_cause_2011, guo_complexity_2013}.

To formulate Max-Cut as a binary variable problem, assign to each vertex a label from $\{+1,-1\}$ indicating to which of the two groups it belongs. Since QAOA is a minimization problem, we use as cost function the (additive) inverse of the number of cut edges. The quantum version of such a cost function can be expressed as a QUBO formula plus a constant term:
\begin{equation}
\label{eq:maxcut-cost-function}
    \hat{C} = -\frac{M}{2} \hat{I} + \frac{1}{2} \sum_{(i,j) \in E} \, \hat{Z}_i \, \hat{Z}_j \; ,
\end{equation}
where $E$ is the set of edges of the graph to partition, $M$ the number of edges, and $\hat{I}$ the identity operator.

\section{Methods}
\label{sec:methods}

A central quantity in the optimization of QAOA circuits is the expectation value of the observable cost function $\hat{C}$ on state $\ket{\gamma,\beta}$ given in \eqref{eq:variational-state}:
\begin{equation}
    F_p (\gamma, \beta) = \bra{\gb} \hat{C} \ket{\gb} \; .
\end{equation}
The minimization of $F_p$ is a good proxy for the desired goal of preparing the ground state of $\hat{C}$. In fact, the ground state is the argument minimizing:
\begin{equation}
    \min_{\ket{\psi}} \bra{\psi} \hat{C} \ket{\psi}
\end{equation}
and it is reasonable to expect that, for expressive enough QAOA circuits, state $\ket{\gb}$ can achieve large overlap with the ground state by minimizing $F_p$ with respect to $(\gb)$.

In practical implementations of QAOA there are two subtle points related to the choice of $F_p (\gb)$ as the function to minimize. First, one does not experimentally access an observable's expectation value directly, but can only estimate it by obtaining many assignments of the problem's variables and averaging their cost function values. This procedure is very natural since $\hat{C}$ is diagonal in the computational basis and the assignments are obtained by measuring all qubits in the Z basis. The expectation value is, however, only approximated by the average and its uncertainty is inversely proportional to the square root of the number of samples. Second, while $F_p$ is the objective function of the minimization, the result reported at the end of QAOA experiments is the minimum cost function value among all the observed samples and not their average.

\subsection{Sampling of Approximation Ratios}
\label{sec:methods-sampling}

To clarify the above remarks, we introduce the ``cost distribution'' associated to $\ket{\gb}$. The QAOA state of $N$ qubits can be represented in the Z basis as:
\begin{equation}
\label{eq:state-representation}
    \ket{\gb} = \sum_{z\in\{-1,+1\}^N} \alpha_z \, \ket{z}
\end{equation}
where $z=(z_1,z_2,\dots,z_N)$ and $\ket{z}$ is the unique common eigenstate of every $\hat{Z}_k$ with eigenvalue $z_k$. Since the cost function is diagonal in the Z basis, one has:
\begin{equation}
    \hat{C} \ket{z} = C(z) \ket{z}
\; .
\end{equation}
The cost distribution of $\ket{\gb}$ is the probability distribution $P_\text{Cost}^{(\gb)}$ that associates to each cost function value $c$ the probability:
\begin{equation}
    P_\text{Cost}^{(\gb)}(c) = \sum_{z \text{ s.t. } C(z)=c} | \alpha_z |^2 \; .
\end{equation}
When clear from the context, superscript $(\gb)$ will be omitted. In the specific case of Max-Cut, we sometimes use the term ``cut distribution'' to indicate the same concept but w.r.t. the cut value. Since the cut value is the additive inverse of the cost value, one has $P_\text{Cut}(c) = P_\text{Cost}(-c)$.

\begin{figure}
    \begin{center}\adjustimage{max size={\linewidth}{0.4\paperheight}}{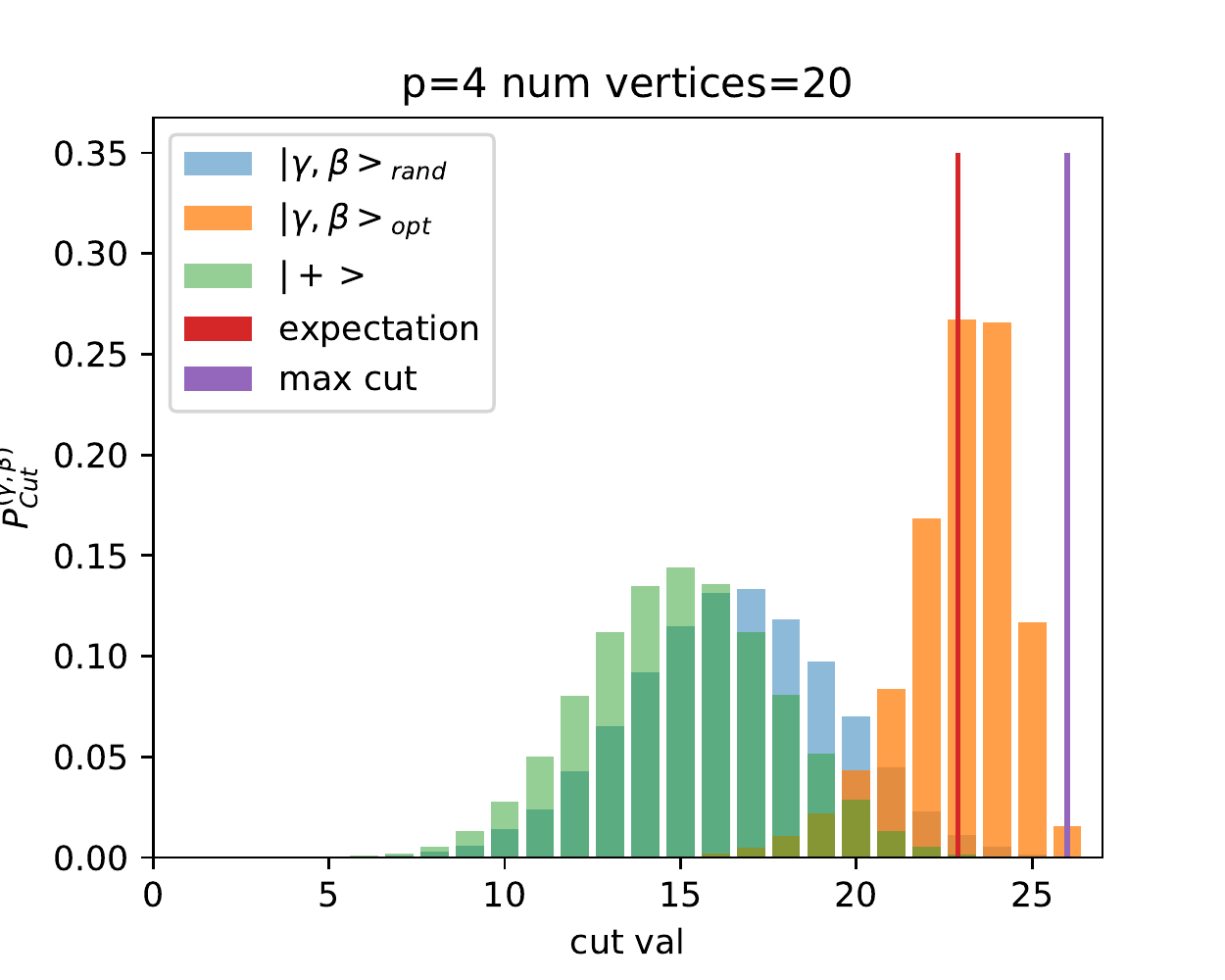}\end{center}
    \caption{Cut distributions for Max-Cut on a random 3-regular graphs with 20 vertices. Three distributions are plotted, each characterized by different values of the parameters $(\gb)$. In green, the case of $\gamma_k = \beta_k =0$ for each $k=0,1,\dots p$ corresponding to the case in which every graph partition has the same probability $1/2^N$. In blue, the distribution for a random choice of parameters, and in orange the one at the end of the optimization process. The vertical red line indicates the expectation value of the orange distribution. The maximum cut value is indicated by the purple column.}
    \label{fig:training-nfev}
\end{figure}

Note that for Max-Cut the probability distribution $P_\text{Cut}$ can be fully described by an exponentially smaller number of values than the $2^N$ amplitudes necessary to describe $\ket{\gb}$. In fact, there are at most $\mathcal{O}(N^2)$ edges in any graph with $N$ vertices, and therefore only the same number of distinct values of $C(z)$. This allows for an effective visualization of the result of optimizing the QAOA circuit. Fig.~\ref{fig:training-nfev} shows the cut distribution when each partition is equally probable (corresponding to a QAOA circuit with $\gamma_k = \beta_k =0$ for each $k=0,1,\dots p$), and compare it to the cost distribution for a random choice of parameters (here $p=4$) and for optimized values of $(\gb)$.

The cut distribution is strictly connected to the result of QAOA experiments. In fact, for a given $(\gb)$, a single experimental run of the QAOA circuit returns a specific graph partition whose cut value is distributed according to $P_\text{Cut}^{(\gb)}$. By repeating the experiment $S$ times and collecting the corresponding cut values, one can estimate $F_p(\gb)$ as their average. Denoting with $z^{(s)}$ the partition observed in the $s$-th run, one has:
\begin{equation}
\label{eq:f_p}
    F_p(\gb) = \lim_{S\rightarrow \infty} \sum_{s=1}^S C \big( z^{(s)} \big) \; .
\end{equation}
In practice, the estimate involves a finite number $S$ of repetitions and past experiments  used $S\sim 1000 - 10,000$ \cite{otterbach_unsupervised_2017, kandala_hardware-efficient_2017, pagano_quantum_2020,arute_quantum_2020}.

Finally, it is important to remember that QAOA is an approximate algorithm which is not guaranteed to return the global solution, but aims at a good approximation. The quality of the solution is expressed in terms of the approximation ratio given by $F_p(\gamma^\prime,\beta^\prime)/c^\star$, where $(\gamma^\prime,\beta^\prime)$ indicates the parameters at the end of the optimization and $C^\star=\min_z C(z)$ is the global minimum of the cost function. However, we highlight that the result of QAOA experiments for Max-Cut should be the largest cut value among all the observed samples and not its average over $P_\text{Cut}^{(\gamma^\prime,\beta^\prime)}$.

For this reason, we structure our analysis of the QAOA performance in terms of the number of circuit repetitions before a cut value above a certain approximation ratio is observed. Formally, consider the approximation ratio $r$ and ask what is the probability of observing at least one cut value above $r\,|c^\star|$ among the first $K$ samples. When such probability exceeds 50\%, meaning that the event is more probable than not, we return $K$ as the expected number of repetitions to reach the desired approximation.

Denoting with $Q_\text{Cut}^{(\gb)}$ the cumulative distribution of $P_\text{Cut}^{(\gb)}$, the probability of observing at least one cut value with approximation ratio at least $r$ while keeping the QAOA parameter unchanged is:
\begin{equation}\label{eq:expected}
    P_r (K) = 1 - \left[ Q_\text{Cut}^{(\gb)}(r\,|c^\star|) \right]^K \,
\end{equation}
since $Q_\text{Cut}^{(\gb)}(r\,|c^\star|)$ represents the probability of \emph{not} observing a cut value above approximation ratio $r$. The expression can be easily adapted to the case of varying $(\gb)$ by considering the actual $Q_\text{Cut}^{(\gb)}$ for each of the $K$ samples. In our approach, we vary $(\gb)$ every $S$ samples (the same parameter as in the finite version of the right hand side of \eqref{eq:f_p}).

\subsection{Quantum Simulation and Circuit Execution Time}
\label{sec:methods-simulation}

All numerical simulations have been implemented using the Python interface of Intel Quantum Simulator (IQS) \cite{guerreschi_intel_2020}. The simulator has been developed in C language for high-performance-computing environments and is released open-source. It stores the state of $N$ qubits as a vector with $2^N$ complex amplitudes that is updated to reflect the execution of quantum operations. The software includes special functionalities to facilitate the emulation of QAOA circuits: instead of decomposing $e^{-i \gamma \hat{C}}$ into one- and two-qubit gates and simulating each of them sequentially, IQS performs a single update that reflects the global operation. In addition, it gives access to the probability distribution of the cut values, namely $P_\text{Cut}^{(\gb)}$ introduced in the previous subsection.

In this numerical study, we have not considered the effect of incoherent errors and environmental noise. The decision was motivated by the substantial overhead required to describe the noise effects, overhead estimated to be a factor $\sim 1,000$. For the considerations of this work, we expect that the noise modifies the probability distribution $P_\text{Cut}^{(\gb)}$ in two ways: by increasing its spread and, for good values of the parameters $(\gb)$, reducing its average cut-value. We leave this point to further investigations. Xue et al.\cite{xue_effects_2019} demonstrated that with local noise the general landscape does not change much, though gradients become flatter (which was further confirmed in Ref \cite{marshall_characterizing_2020}). It was also recently shown on the Google Sycamore device that the QAOA landscape for the Sherrington-Kirkpatrick model is similar to the simulated one \cite{arute_quantum_2020}. More work is necessary to address how local noise causes deviations in the expected cost function and what the relative trade-offs are with respect to system size, circuit depth and noise rate.

In the following analysis, we do not need to report the time to simulate QAOA circuits, but rather estimate the time to execute them on actual hardware. In addition, we are not interested in modelling a specific quantum processor, but only in providing an order of magnitude estimate.
To this extent, we consider two abstract architectures characterized by the same gate set but different connectivity: all-to-all and  bi-dimensional square grid. The gate set includes parametrized one-qubit and two-qubit rotations, like $\exp{(-i \beta \hat{X}_k)}$ and $\exp{(-i \gamma \hat{Z}_j \hat{Z}_k)}$, and all gates have the same duration  $T_\text{gate} = 10\,$ns. We assume additional $T_{spam} = 1\,\mu$s for state preparation and measurement (SPAM). These values are reasonable in the context of emerging NISQ-era architectures, especially for superconducting transmon qubits \cite{arute_quantum_2019}. No further limitations of gate parallelism are considered, apart from the requirement that every qubit can be involved in at most one gate at a time. For the circuit compilation and to satisfy the connectivity constraints, we adopt the scheduler based on dynamical pattern optimization described in reference \cite{guerreschi_scheduler_2019}.

The two architectures give different estimates of the execution time for a single QAOA circuit. For the full connectivity, given a graph instance with maximum vertex degree $d$, one has:
\begin{equation}
\label{eq:walltime_fc}
    T_\text{circuit}^{(FC)} = T_\text{spam} + ((d+1) T_\text{gate} + T_\text{gate} ) p \; ,
\end{equation}
where we used the result that one needs at most $(d+1)$ colors to color the edges of a graph with vertex degree $d$ \cite{bondy_graph_1976} and therefore $(d+1)$ layers of two-qubit gates may be required to implement each $e^{-i \gamma_k \hat C}$, while a single layer suffices for each $e^{-i \beta_k \hat B}$. We do not consider extra time between circuit repetitions in addition to $T_\text{spam}$, not even when $(\gb)$ needs to be updated (once in $S$ repetitions).

For the square grid connectivity, the scheduler returns the circuit depth $l$ in terms of the number of layers of parallel gates. The value of $l$ varies from instance to instance. The time to execute a single QAOA circuit is:
\begin{equation}
\label{eq:walltime_2d}
    T_\text{circuit}^{(2D)} = T_\text{spam} + l \, p \, T_\text{gate} \; .
\end{equation}

\subsection{Optimization of Variational Parameters}
\label{sec:methods-optimization}

The performance of QAOA depends critically on the choice of depth $p$ and variational parameters $(\gb)$. Due to its connection with adiabatic quantum optimization \cite{mbeng_quantum_2019}, QAOA is capable of finding the global minimum in the ${p\to\infty}$ limit \cite{farhi_quantum_2017}, but its performance is still being explored for finite $p$ \cite{hastings_classical_2019,bravyi_classical_2019,bravyi_obstacles_2019,farhi_quantum_2020,farhi_quantum_2019}. Early results suggested that for $p=1$ QAOA provides an approximation ratio for the problem MAX3Lin2 that is better than its classical counterparts. However the improvement has since been overtaken by a new classical method \cite{barak_beating_2015}.

A number of strategies have been utilized in recent work to determine optimized variational parameters, including local optimization \cite{shaydulin_evaluating_2019}, gradient-based methods \cite{guerreschi_practical_2017}, multi-start optimization \cite{shaydulin_multistart_2019}, reinforcement learning \cite{khairy_learning_2019, wauters_reinforcement_2020}, and analytical prediction \cite{streif_training_2019, farhi_quantum_2019}. In this work, we utilize a combination of local and multi-start optimization, and discuss the results from each.  We utilize the Asynchronously Parallel Optimization Solver for finding Multiple Minima (APOSMM) \cite{larson_asynchronously_2018}, which has been used successfully with QAOA and graph clustering \cite{shaydulin_multistart_2019}. For local optimization we use the derivative-free Bound Optimization By Quadratic Approximation (BOBYQA) \cite{powell_bobyqa_nodate}. For APOSMM, we vary the total number of evaluations of the cost function $n_\text{fev}$ as discussed in Section~\ref{sec:results}.

\begin{figure}
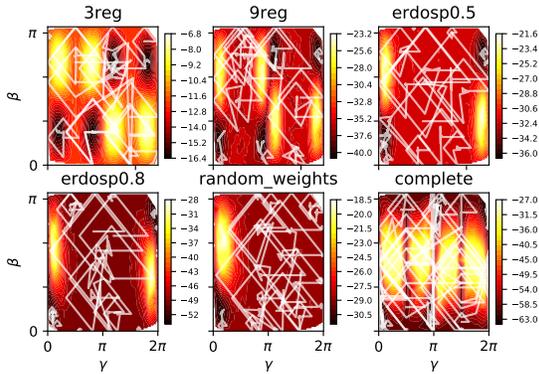

    \begin{center}\adjustimage{max size={\linewidth}{0.5\paperheight}}{fig-contour.pdf}\end{center}
    \caption{Optimization landscape (see \eqref{eq:cost-function}) for QAOA applied to the Max-Cut problem on graphs with $N=16$ vertices for p = 1. Several type of graphs are considered, from left to right and top to bottom: random 3-regular, random 9-regular, Erd\H{o}s $p_E=0.5$, Erd\H{o}s $p_E=0.8$, random weights, and the complete graph. As the graph structure changes and noticeably as the vertex degree increases, first from random 3- to 9-regular and then to Erd\H{o}s and random weights, the global minimum becomes narrower and local plateaus appear. Shown in white are traces from the different local optimization attempted by APOSMM in its search for the global minimum.
    }
    \label{fig:APOSMM}
\end{figure}

To demonstrate the APOSMM optimization process, \figref{fig:APOSMM} shows the path followed by the optimizer in the parameter landscape with $p=1$. In this work, we consider graphs of three types:
\begin{description}
    \item[random $k$-regular]: each and every vertex has exactly $k$ edges, chosen randomly;
    \item[Erd\H{o}s-Renyi]: each of the $N(N-1)$ edges has probability $p_E$ of being present, with unbounded vertex degree
    \item[random weights]: complete graph with weights chosen uniformly at random in $[0,1]$ (see Appendix \ref{sec:appendx-random} for a note on random weight graphs). 
\end{description}
Each type can give rise to a distinct graph class, for example by specifying the non-negative integer parameter $k$ or the continuous parameter $p_E\in[0,1]$.

For each graph type, the optimization landscape is distinct.  For 3-regular graphs, the landscape is characterized by minima separated without plateaus. The variational parameters exhibit periodicity within the domain considered, which has been demonstrated previously \cite{zhou_quantum_2018, wang_quantum_2018}. For Erd\H{o}s and random weights, the landscape minima begin to concentrate and plateaus appear.  These plateaus include many local minima that are suboptimal and act as trap for the local optimization making it challenging for APOSMM to find the global minimum (the path in parameter space is drawn as white traces). Similar difficulties in finding optimized parameters have been noted for QAOA and associated to the concentration of optimal parameters into a small region of the parameter space for random weight graphs \cite{farhi_quantum_2019}.

\begin{figure}
    \begin{center}\adjustimage{max size={\linewidth}{0.5\paperheight}}{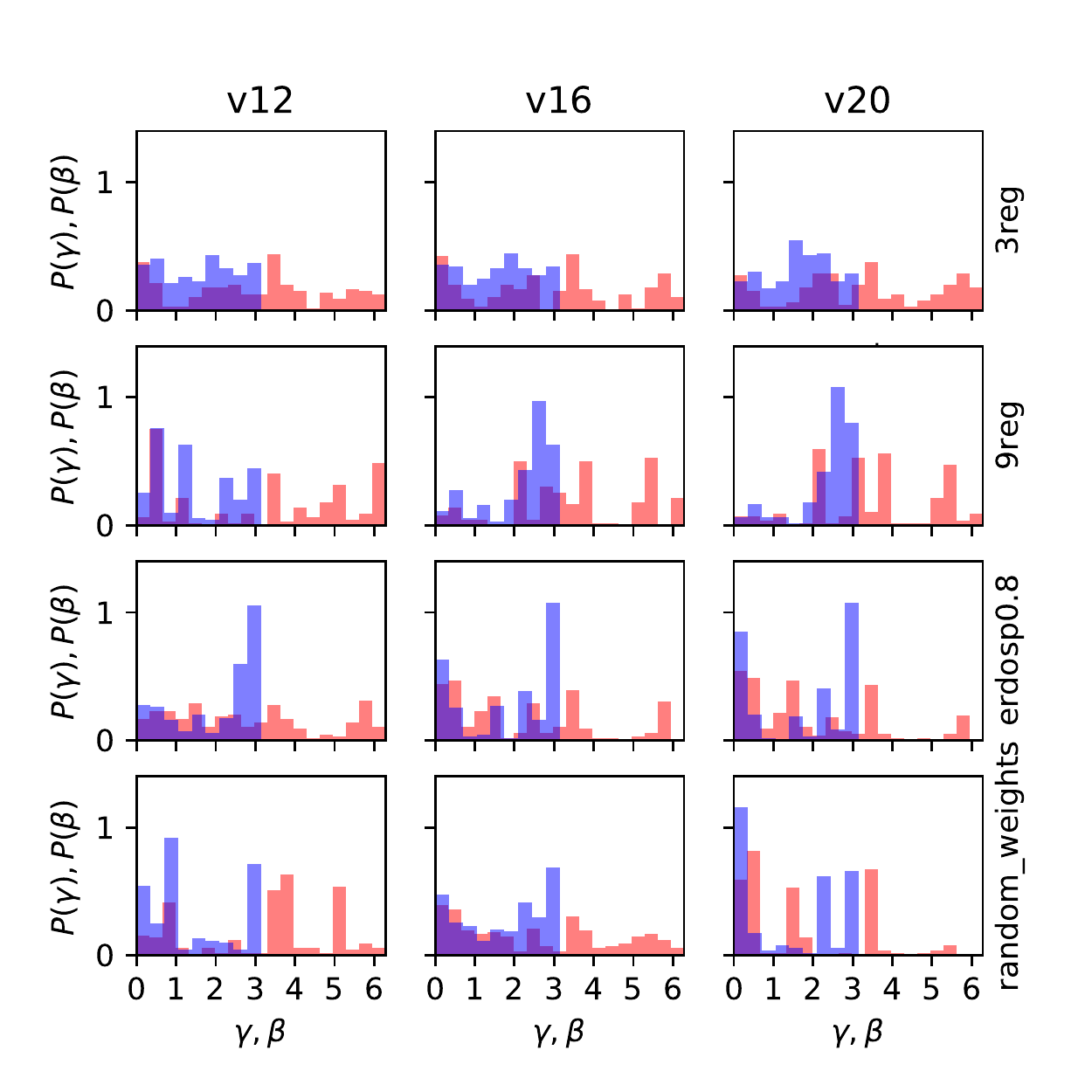}\end{center}
    \caption{Probability density of variational parameters $P(\gamma)$, $P(\beta)$ with $p=4$ and for different graph types and sizes. (b) Overall difference in the probability densities   $|P_a(\gamma)-P_b(\gamma)|$ for varying nnodes $a$ and $b$.  For all graph types there is a reduction in the difference with increasing nnodes, although minimal for random weights (see Appendix Figure \ref{fig:concentration-diff}).
    }
    \label{fig:concentration}
\end{figure}

Recently, it has been observed that good parameters obtained after the optimization of a specific instance, perform well also for other instances of the same class and even for different instance sizes. Analytical arguments \cite{brandao_for_2018} suggest that the numerical observations reported in \cite{barkoutsos_improving_2019, shaydulin_evaluating_2019} may apply to a broad class of random graphs. We collect supporting evidence in \figref{fig:concentration}, where a partial concentration of the optimized parameter values is suggested.
There are two aspects worth observing: first, in any single plot the probability distribution has a structure connected to the concentration of optimized values of the parameters. Second, for each graph type, moving along the same row from left to right (\emph{i.e.} increasing $N$), the distribution seems to converge suggesting that it may be preserved for $N>20$.
We elaborate more on the latter observation in Appendix Figure \ref{fig:concentration-diff}, but mention that the effect is evident for random 3-regular graphs, but less marked for Erd\H{o}s and random weight graphs.

In our case, \figref{fig:concentration} is hardly conclusive. However, in the next Section we utilize the ``trained'' parameter sets optimized from small instance sizes to predict the performance for larger sizes. The effectiveness of this technique can be seen as \textit{a posteriori} justification of this choice.
\begin{figure*}
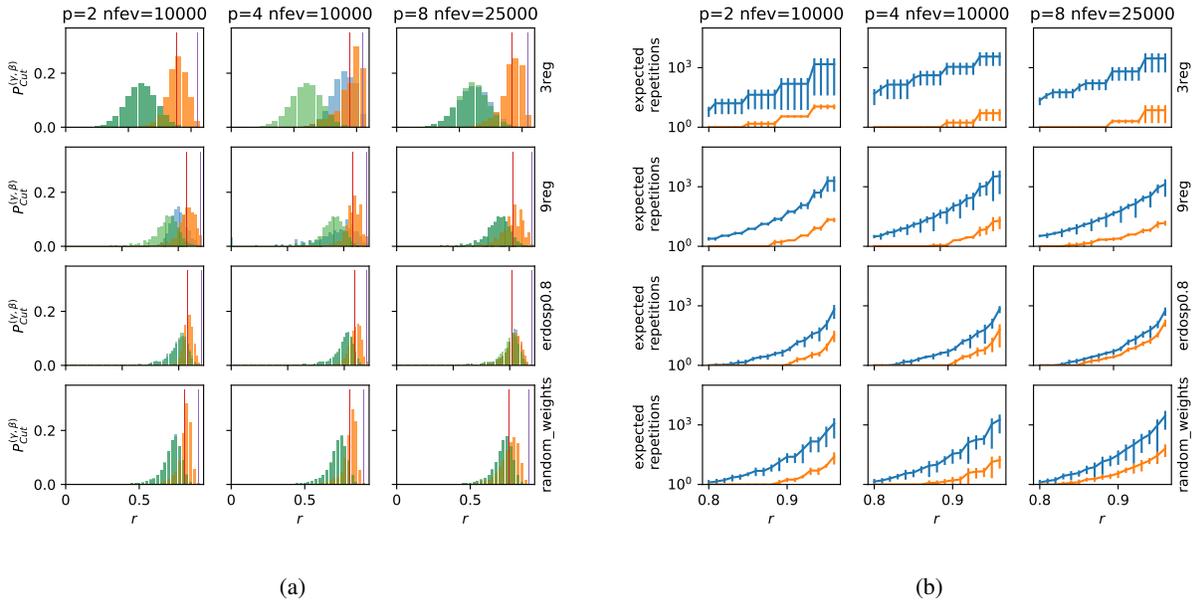

\centering
\begin{subfigure}{1.0\columnwidth}
\adjustimage{max size={\linewidth}{0.4\paperheight}}{fig-training-p.pdf}
\caption{}%
\label{subfiga}%
\end{subfigure}\hfill%
\begin{subfigure}{1.0\columnwidth}
\adjustimage{max size={\linewidth}{0.4\paperheight}}{fig-training-v16-expected.pdf}
\caption{}%
\label{subfigb}%
\end{subfigure}\hfill%
\caption{
(a) Cut distributions for Max-Cut and several graph types, for a typical instance with $N=16$ vertices (note the axes are scaled to approximation ratio $r$). From top to bottom: random 3-regular, random 9-regular, Erd\H{o}s $p_E=0.8$, and random weight graphs. In the first column, $p=2$ and APOSMM is given a budget of $n_\text{fev}=10,000$ function evaluations; second column $p=4$ and $n_\text{fev}=10,000$; third column $p=8$ and $n_\text{fev}=25,000$.
Three distributions are plotted, each characterized by different values of the $2p$ parameters. In green, the case of $\gamma_k = \beta_k =0$ for each $k=1,2,\dots p$ corresponding to the case in which every graph partition has the same probability $1/2^N$. In blue, the distribution at an intermediate stage of the optimization and in orange the distribution for the optimized parameters. Note the data colors and vertical lines are the same as in Figure \ref{fig:training-nfev}
(b) Expected number of repetitions of the QAOA circuit required to observe at least one samples with given approximation ratio $r$. The blue line is obtained updating $P_\text{Cut}$ every $S=1,000$ samples during the optimization process. The orange line corresponds to sampling directly from the optimized distribution. We report the average over 10 instances and the vertical bars represent one standard deviation. Subplots are organized in rows and columns according to panel (a).}
\label{fig:training-expected-nfev-p}
\end{figure*}

\section{Results}
\label{sec:results}

We visualize the cut probability $P_\text{Cut}$ for several graph types and three $p$ values in \figref{fig:training-expected-nfev-p}~(a). The green distribution corresponds to the $p=0$ case, while the orange distribution to $p=2,4,8$ ---from left to right--- and optimized parameters $(\gb)$. Panel (b) of the same figure reports the cost of parameter optimization. The optimization budget of APOSMM is denoted by $n_\text{fev}$ and varies from 10,000 to 25,000 function evaluations. Each function evaluation corresponds to an estimate of $F_p$ with $S=1,000$ samples, a number smaller than those used in QAOA experiments to date \cite{otterbach_unsupervised_2017, kandala_hardware-efficient_2017, arute_quantum_2019, arute_quantum_2020,jurcevic_demonstration_2020}. For 3- and 9-regular graphs, optimized variational parameters are found for $p=4,8$ yielding distributions which have large overlap with the maximum cut value (see also Figure \ref{fig:training-nfev}). For Erd\H{o}s and random weight graphs, the distributions have less overlap, and increasing $p$ and $n_{\text{fev}}$ does not lead to an increase in overlap. 

This can be seen in Figure \ref{fig:training-expected-nfev-p}~(b), which shows the expected number of circuit's repetitions before the probability of observing a cut value with a given approximation ratio is at least 50\% (see equation \eqref{eq:expected}). The blue line corresponds to the case in which the parameters are randomly initialized and then updated during the optimization. The orange line represents the case in which the optimized parameters are known from the start. It is not surprising that this knowledge reduces the QAOA cost by several orders of magnitude. The results in this panel are averaged over 10 random instances of each graph type, whereas panel (a) shows a typical instance.

For $k$-regular graphs and small instance size, shallow circuits corresponding to $p\leq8$ are expressive enough for state $\ket{\gb}$ to exhibit a large overlap with the ground state of $\hat{C}$. This is not the case for Erd\H{o}s and random weights, resulting in over an order of magnitude larger expected number of circuit repetitions for a given approximation ratio. We project the implications of these results in the next figure by comparing the performance with classical exact and approximate solvers.

\begin{figure}
    \begin{center}\adjustimage{max size={\linewidth}{0.4\paperheight}}{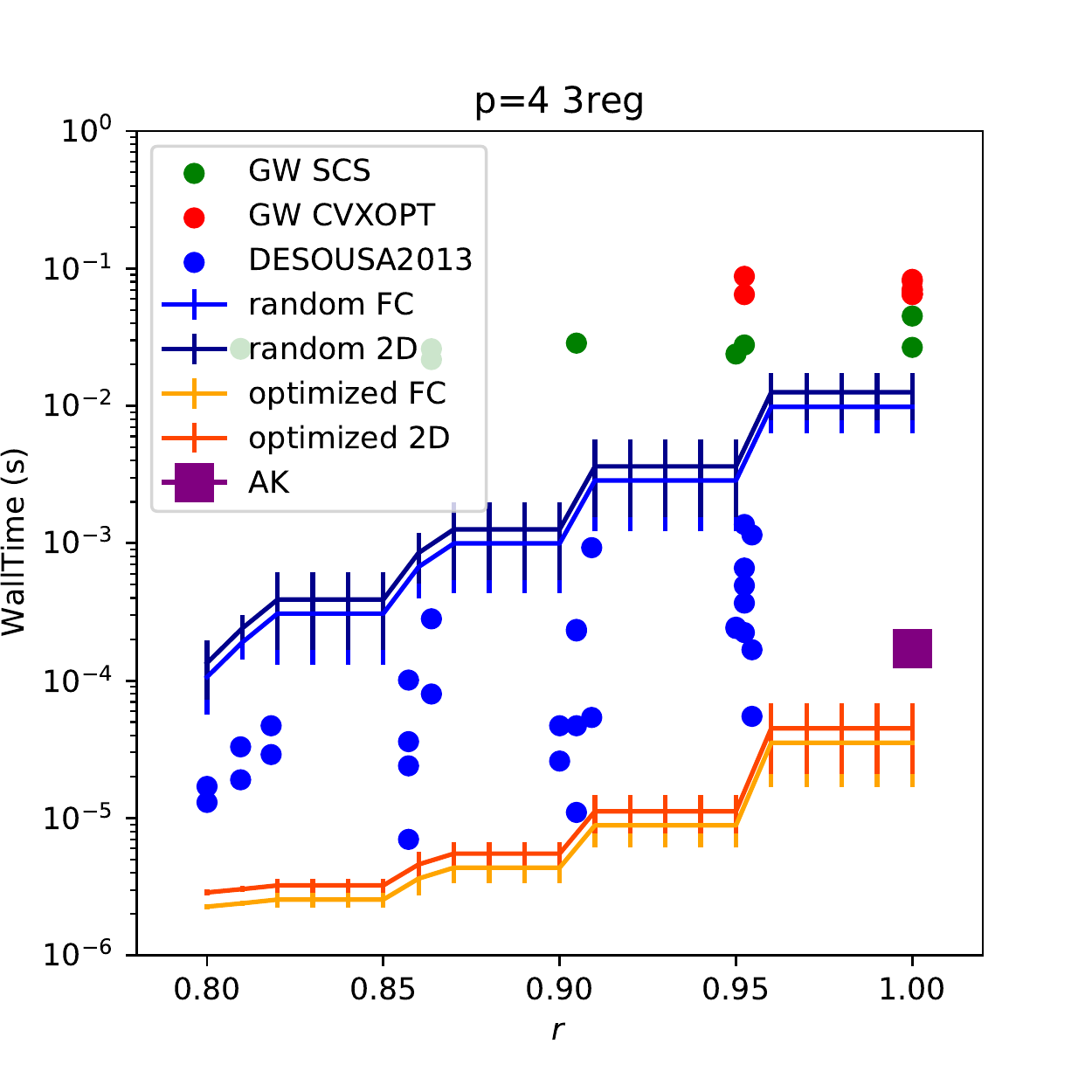}\end{center}
    \caption{The WallTime performance for random 3-regular graphs with 16 vertices, averaged over 10 instances. The horizontal axis indicates the desired quality of the solution, expressed in terms of the approximation ratio $r$. The blue lines correspond to the performance of the local optimization attempts of the APOSMM training stage, with a large $n_\text{fev}$ budget and initial random parameters. The label FC indicates results for a quantum processor with full connectivity and 2D with a square grid connectivity. The orange lines correspond to the performance using the optimized parameters found by training. For comparison, classical exact solvers (AKMAXSAT) and approximate algorithms (GW) are also shown. Two semi-definite programming implementations (labelled SCS and CVXOPT from the name of their Python package) are used for GW. Vertical and horizontal (for GW) bars indicate one standard deviation of the instance average. More considerations in the main text.}
    \label{fig:walltime-example}
\end{figure}

We utilize the simple model for quantum processors described in Section~\ref{sec:methods-simulation}: noiseless devices with full and bi-dimensional square grid connectivity. The time to run a single QAOA circuit in the two architectures is given by \eqref{eq:walltime_fc} and \eqref{eq:walltime_2d} respectively. When multiplied by the number of circuit repetitions, we obtain the absolute-time performance of QAOA, here called WallTime. Figure \ref{fig:walltime-example} shows the WallTime of QAOA averaged over 10 random 3-regular graphs, for both quantum architectures and starting from random and optimized $(\gb)$ values. Also shown is the performance for a state-of-the-art exact solver (AKMAXSAT) \cite{kuegel_improved_2012} and a famous approximate algorithm (GW) \cite{goemans_improved_1995}.
We report the time to solution on the vertical axis, together with the quality of the approximate solution, expressed by its approximation ration $r$. It is clear that the exact solver provides a single datapoint located at $r=1$, while the approximate solvers lead to a distribution of results. In particular, we observe that the approximation ratio performance varies significantly for GW depending on which optimization method (CVXOPT or SCS) is used to solve its semi-definite relaxation. The performance of random and optimized QAOA varies by over 2 orders of magnitude, with optimized QAOA producing approximate and exact solutions competitively with both GW and AKMAXSAT.

We observe that the relative performance of GW compared to AKMAXSAT is surprising: an approximate solver is expected to be faster than an exact solver. Clearly this is not the case in our numerical study. There are multiple reasons for this situation and we discuss them in Appendix~\ref{sec:appendx-classical}. Here we mention that the advantage of heuristic algorithms is more relevant for large $N$, where their better scaling takes over small size effects and certain details of the implementation. For these system sizes, the approximation ratio achieved by GW is more relevant to the performance comparison. 

\begin{figure}
    \begin{center}\adjustimage{max size={\linewidth}{0.4\paperheight}}{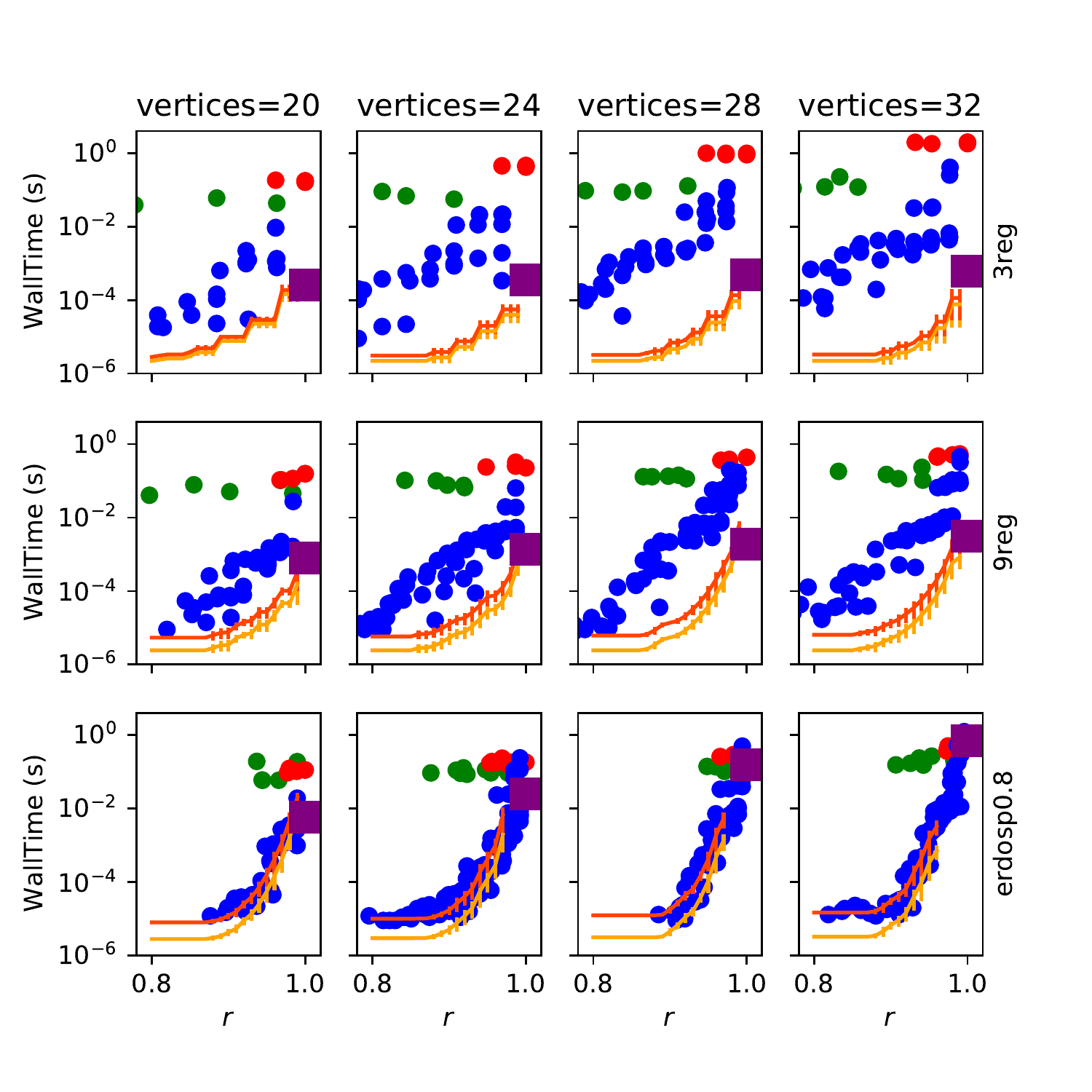}\end{center}
    \caption{WallTime performance for QAOA, utilizing the trained parameters from the 16 vertex graphs (see Fig. \ref{fig:walltime-example}), and compared against the classical exact (AKMAXSAT) and approximate (GW) solvers. For all graph types, QAOA is able to produce approximate solutions which are competitive with GW SCS and CVXOPT. For exact solutions, QAOA is also competitive, but with evidence of an exponential dependence on $r$, especially for  Erd\H{o}s.  Results for the classical solvers for these instance sizes should be taken in the context of the asymptotic performance noted in Appendix \ref{sec:appendx-classical}.   
     }
    \label{fig:walltime-average}
\end{figure}

Figure~\ref{fig:walltime-average} shows the WallTime performance of the optimized variational parameters from Figure~\ref{fig:training-expected-nfev-p} for different graph types and graph sizes up to $N=32$ We use the set of trained parameters from $N=16$ (one set per instance) and average over these. Here we take advantage of the recent observation that good parameter values are concentrated and shared among instances of the same type. We empirically verify that good parameter values can be applied to instances with more vertices. By offloading the cost of the parameter optimization, the WallTime performance of QAOA for 3- and 9-regular graphs remains competitive with classical solvers. We observe an exponential increase of the WallTime before reaching approximation ratio close to 1, indicating an exponentially smaller value of $P_\text{Cut}(|c^\star|)$ compared to, for example, $P_\text{Cut}(0.9 \, |c^\star|)$. This behavior is very clear for Erd\H{o}s graphs. For 3- and 9-regular, there is some evidence of linear$/$sublinear behavior in WallTime versus approximation ratio, another indication of the effective overlap of the distributions shown in Figures \ref{fig:training-nfev} and \ref{fig:training-expected-nfev-p} with the maximum cut. 

\section{Discussion}
\label{sec:Discussion}

The results discussed in Sections~\ref{sec:methods} and \ref{sec:results} demonstrate how much QAOA's performance depends on the graph type and optimization method. There is a fundamental performance gap between 3- and 9-regular and the Erd\H{o}s and random weights graph types. The transformation of the optimization landscape (\figref{fig:APOSMM} and, indirectly, \figref{fig:concentration}) results in the need for a larger optimization budget ($n_\text{fev}>25,000$) and an inability for the $p$ considered to produce states $\ket{\gb}$ with large overlap with the maximum cut for Erd\H{o}s and random weight graph types. In this context, a very practical question is how to adapt the QAOA protocol to specific graph types and, more generally, to the class of problem instances.

The strong performance dependence on graph type is also true for classical exact and approximate solvers. For example, we observe a correlation between the time required by AKMAXSAT and the vertex degree of the graph. This is confirmed by the trivial case of fully connected graphs, in which any balanced assignment with $N/2$ vertices in both partitions is a global solution: AKMAXSAT requires a time exponential with $N$ and with a much larger exponent than any other graph type.

A central observation of our study is that, by utilizing the parameters optimized from small instance sizes, the WallTime for large instance sizes is dramatically reduced. The parameters' optimization can be compared to a training process and its cost off-loaded to a preliminary phase. This is the difference between the blue lines of \figref{fig:walltime-example} where no training is assumed, to the orange lines in \figref{fig:walltime-example} and \figref{fig:walltime-average} where the sampling starts from pre-trained parameters. This additional cost in terms of circuit evaluations is up to $n_\text{fev}\,S = 2.5\times 10^7$ 
and, when converted to WallTime, is on the order of 10--100~s. This is the same order of magnitude reported in reference~\cite{guerreschi_qaoa_2019}. 

As a consequence, the WallTime performance reported in \figref{fig:walltime-average} depends critically on minimizing the training process by utilizing previously-trained parameters. Our results suggest that the effectiveness of this approach varies significantly for the graph types studied, and more work is necessary to assess how this process scales to instance sizes larger than $N \sim 50$. Along the same lines of avoiding the training phase, recent works propose the use of machine learning techniques \cite{khairy_learning_2019, alam_accelerating_2020} or a mix of analytical and numerical results at $N\rightarrow\infty$ \cite{streif_training_2019} to identify good parameters.

Finally, we comment on the choice of the number of QAOA steps. In this work, we have fixed $p$ at the beginning of the optimization. However, there is growing evidence of the need for $p$ increasing with  instance size $N$, without a clear understanding of the exact dependence \cite{bravyi_classical_2019,bravyi_obstacles_2019,farhi_quantum_2020}. From a practical perspective, one may want to extend the optimization process with an adaptive search of $p$. For the small graphs considered in this study, \figref{fig:training-expected-nfev-p}~(b) suggests that such overhead is unjustified when the goal is minimizing QAOA's WallTime. In fact, the main difference from $p=2$ to $p=8$ is the reduced variance of the distribution over instances of the same graph type, without a pronounced reduction of the number of circuit repetitions (also, notice that circuits with $p=8$ take longer to execute than circuits with $p=2$). 


\section{Conclusions}
\label{sec:conclusions}

We introduced new performance metrics for characterizing the performance of QAOA, focusing on the probability of observing a sample above a certain quality. In this work, given a desired approximation ratio, we formulate the performance of QAOA as the time needed before at least one sample with approximation ratio above the desired threshold is observed with probability at least 50\%.

By combining this new approach with ``training'' of the QAOA variational parameters, our results show a reduction in the QAOA execution time for Max-Cut on random 3-regular graphs of two orders of magnitude with respect to previous estimates \cite{guerreschi_qaoa_2019}. This was possible due to the reduction in number of samples in the calculation of the average approximation ratio
, and because we do not wait for the convergence of the parameters' optimization before analyzing the candidate solutions obtained as single samples.

For other graph types, namely Erd\H{o}s and random weight, the performance of QAOA varies significantly. While sampling-based results show performance improvement for the instance sizes studied, the actual performance test for approximate solution requires its assessment for instance sizes $\sim 100 - 1,000$ depending on the graph type (see Appendix \ref{sec:appendx-classical}). To this end, more work is needed to study how effective the training and sampling can be extended to more challenging instance sizes, and how this approach can be extended beyond fixed $p$ (for example a mild dependence $p \propto log(N)$) while including the effect of noise in the analysis.


\section*{Acknowledgement}

This material is based upon work funded and supported by the Department of Defense under Contract No. FA8702-15-D-0002 with Carnegie Mellon University for the operation of the Software Engineering Institute, a federally funded research and development center. Carnegie Mellon® is registered in the U.S. Patent and Trademark Office by Carnegie Mellon University. DM20-0424

This work used the Extreme Science and Engineering Discovery Environment (XSEDE), which is supported by National Science Foundation grant number ACI-1548562. Specifically, it used the Bridges system, which is supported by NSF award number ACI-1445606, at the Pittsburgh Supercomputing Center (PSC).

\bibliographystyle{unsrt}
\bibliography{sei-quantum}

\newpage

\appendix\label{sec:appendix}

\subsection{Classical Performance and Graph Types}
\label{sec:appendx-classical}

\begin{figure}[b!]
  \centering
  \begin{subfigure}{0.56\textwidth}
    \includegraphics{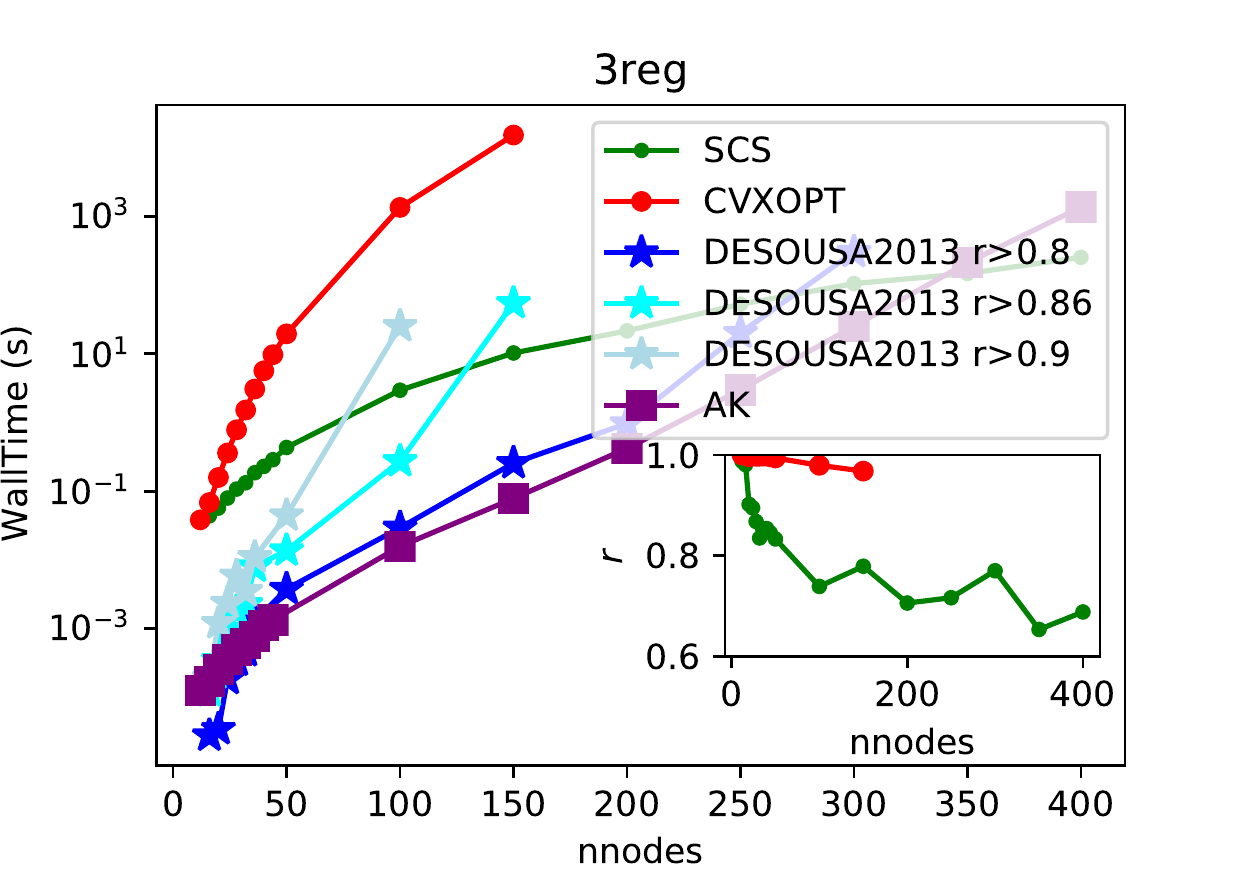}
    \caption{}
    \label{fig:demo1}
  \end{subfigure}
   \begin{subfigure}{0.56\textwidth}
    \includegraphics{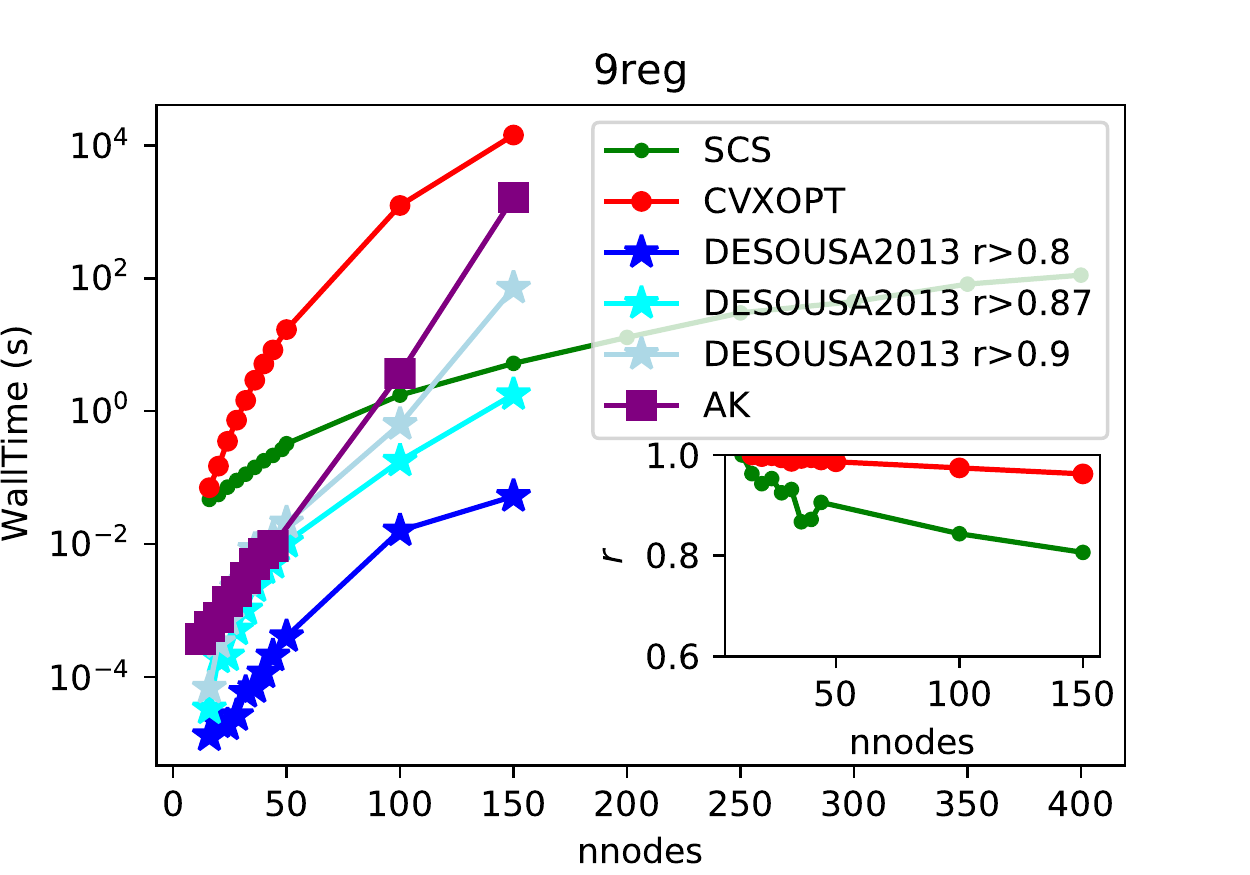}
    \caption{}
    \label{fig:demo2}
  \end{subfigure}
  \caption{WallTime and approximation ratio performance of classical algorithms. We consider the exact solver AKMAXSAT (AK) \cite{kuegel_improved_2012}, the Goemans-Williamson (GW) algorithm \cite{goemans_improved_1995} implemented with either the SCS or the CVXOPT convex optimizer, and the heuristic solver DESOUSA2013 \cite{dunning_what_2018-1} for graph types 3-regular (panel a), 9-regular (panel b). Bounded degree d-regular with fixed d can be approximated to within 0.8785 + $\mathcal{O}(d)$, with 0.921 for $d= 3$ \cite{feige_improved_2002}.   }\label{fig:appendix:dreg}
\end{figure}

Because of the evolving complexity-theoretic assessment of QAOA, a framework for the search for quantum advantage needs to incorporate a large number of problem instances and  all  three  solver  modalities:  exact, approximate,  and  heuristic. Max-Cut is NP-hard to approximate better than 16$/$17 \cite{hastad_optimal_2001,trevisan_gadgets_2000}. It is also APX-complete, and if the Unique Games Conjecture is true \cite{khot_optimal_2007-3}, semidefinite programming and the Goemans and Williamson (GW) random hyperplane-rounding technique achieves the best known approximation $r=0.878567$ \cite{goemans_improved_1995}. In Ref.~\cite{dunning_what_2018}, de Sousa et al. (2013) developed a new Max-Cut heuristic for use in an image segmentation application. This heuristic was chosen for its  performance with the graph types considered here \cite{dunning_what_2018}, based both on the quality of the solution and the time required to reach it.

\begin{figure}[b!]
  \centering
  \begin{subfigure}{0.56\textwidth}
    \includegraphics{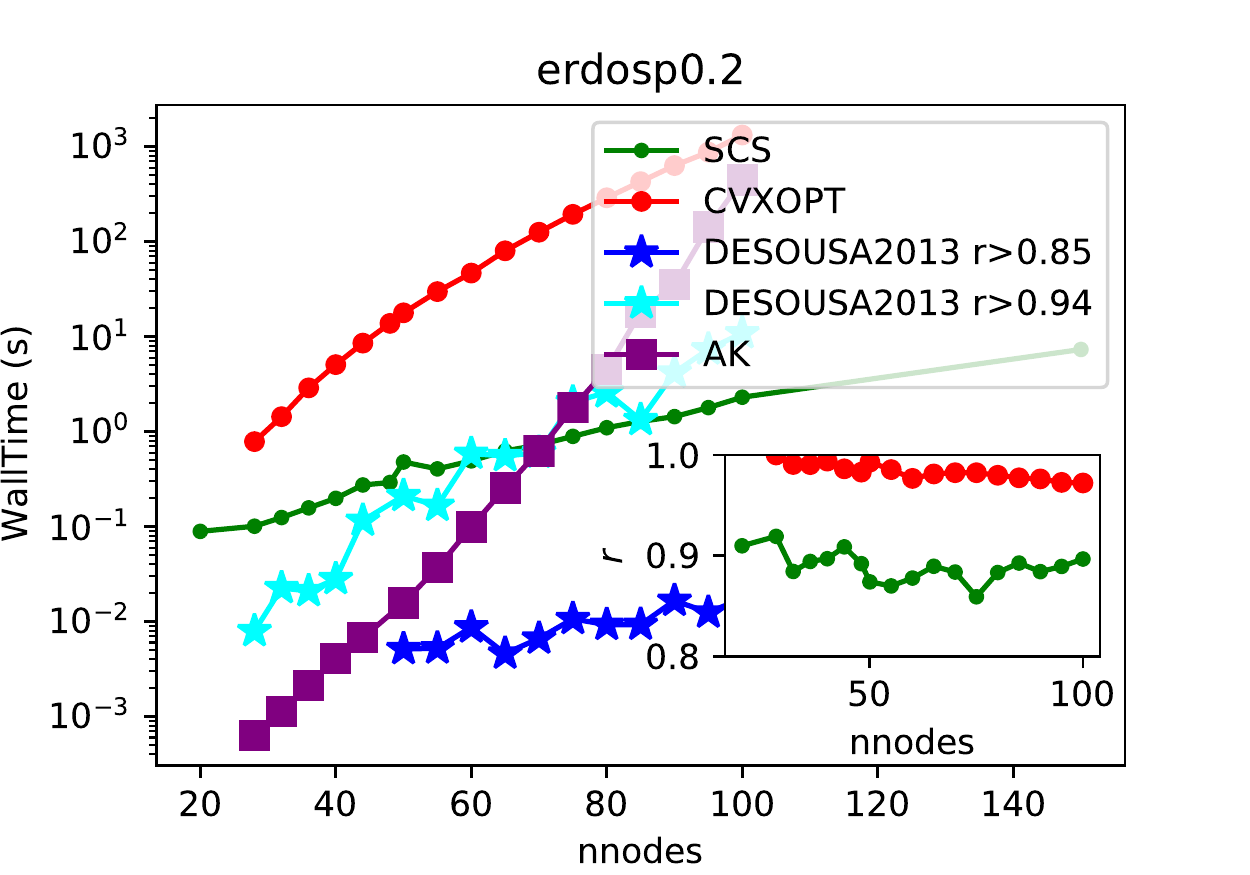}
    \caption{}
    \label{fig:demo3}
  \end{subfigure}
  \begin{subfigure}{0.56\textwidth}
    \includegraphics{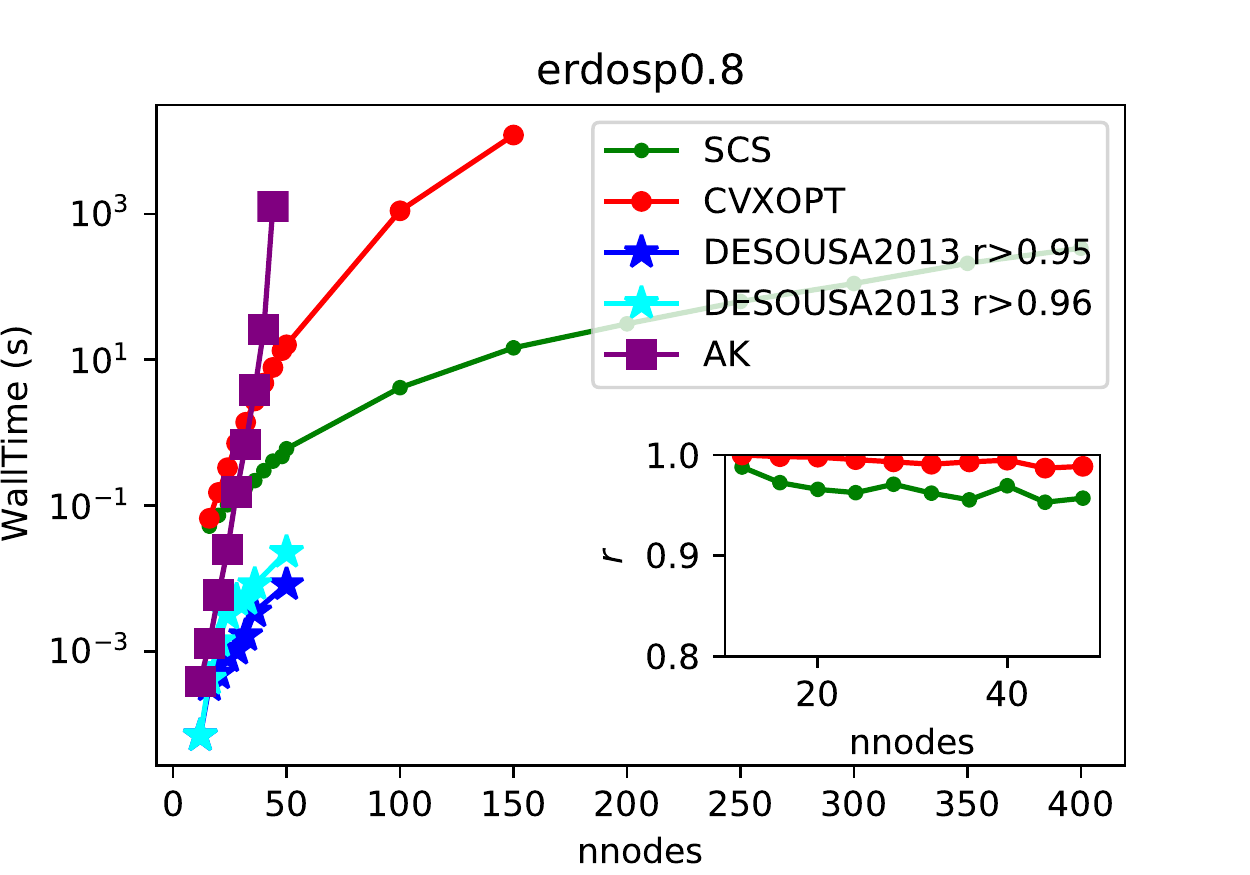}
    \caption{}
    \label{fig:demo5}
  \end{subfigure}    
  \caption{WallTime and approximation ratio performance of classical exact solver AKMAXSAT (AK) \cite{kuegel_improved_2012}, the Goemans-Williamson (GW) algorithm \cite{goemans_improved_1995}, and the DESOUSA2013 heuristic solver \cite{dunning_what_2018-1} for graph type Erd\H{o}s with $p_E=0.2$ (panel a) and $0.8$ (panel b). Max-Cut can be mapped to the Max-2-SAT problem, which undergoes a ``phase'' transition at the ``critical'' point $c=1$ (in the context of Max-Cut, $p_{E}=0.5$) with the expected number of clauses not satisfied by an optimal solution quickly changing from $\mathcal{O}(1/n)$ to $\mathcal{O}(n)$ \cite{coppersmith_random_2004}. }\label{fig:appendix:erdos}
\end{figure}

Performance results are shown in Figures~\ref{fig:appendix:dreg} and \ref{fig:appendix:erdos} for $d$-regular and Erd\H{o}s graph types respectively. For GW, 100 random hyperplanes are used to sample each graph instance.  For all solvers, 10 graph instances are averaged over. The Python CVXOPT library \cite{sra_optimization_2011} implementation of GW used in this study has large interpreted overhead, making its WallTime performance for very small instance sizes less relevant. In general, the GW algorithm is sensitive to the optimizer used \cite{goemans_improved_1995,keetch_max-cut_2017}, and  here we consider SCS and CVXOPT. This results, most pronounced for 3-regular and SCS optimization, in a failure to produce the  approximation guarantee of 0.878567.  Across all graph types SCS under-performs in quality compared to CVXOPT, but runs in faster computational time asymptotically.
The performance of the exact solver AKMAXSAT\cite{kuegel_improved_2012} is overtaken by GW at instance sizes that depend on the graph type, but are outside the scope of the simulations of quantum computing discussed in \secref{sec:methods-simulation}. 
All numerical benchmarks run on the Bridges supercomputer hosted by the Pittsburgh Supercomputing Center (PSC). Each computing node is a two-socket system equipped with Intel Haswell (E5-2695 v3) processor (14 cores per CPU) \cite{noauthor_system_nodate}.

By extending the numerical results to graph sizes (quantified by the number of vertices denoted with ``nnodes'') inaccessible to quantum simulators (see Figure~\ref{fig:walltime-average} of the main text), we want to present the asymptotic difference between exact, approximate, and heuristic solver performance. The choice of graph types in Sections~\ref{sec:results} are influenced by the limited graph sizes available for quantum computing simulation. For quantum advantage, the performance tests require instance sizes $\sim 100 - 1,000$ \cite{dunning_what_2018-1}. 

\subsection{Random Weight Graphs}
\label{sec:appendx-random}

Notice that the last graph type has non-unit edge weights. The Max-Cut problem is extended by counting not the number, but the total weight of the edges cut by the partition. In the literature this is known as the weighted Max-Cut problem. In this case, the cut probability $p_\text{cut}$ is a continuous probability density, and we represent it in discretized form according to $P_\text{cut}(c)=\int_c^{c+1} p_\text{cut}(c^\prime) \text{d}c^\prime$.

\subsection{On the convergence of optimized parameters}
\label{sec:appendx-convergence}

\begin{figure}[tp]
    \begin{center}\adjustimage{max size={\linewidth}{0.5\paperheight}}{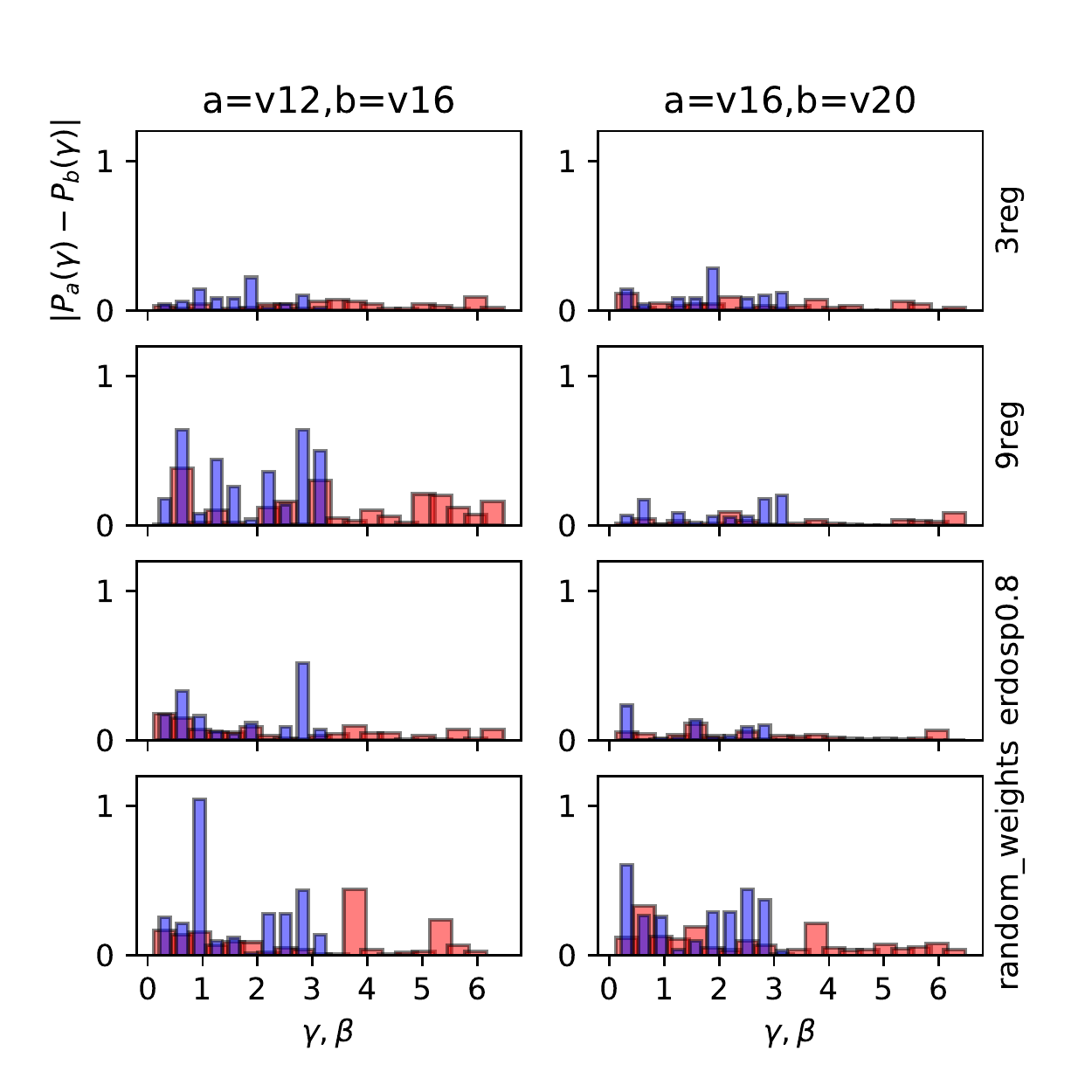}\end{center}
    \caption{Absolute difference in the probability densities   $|P_a(\gamma)-P_b(\gamma)|$ for varying number of vertices $a$ and $b$ computed from the probability densities in \figref{fig:concentration}. For all graph types there is a reduction in the difference with increasing number of nodes, although minimal for random weights.
    }
    \label{fig:concentration-diff}
\end{figure}

As mentioned in the main text, recent works have suggested that QAOA circuits may approximately solve Max-Cut instances of the same class by using the same parameters \cite{brandao_for_2018, barkoutsos_improving_2019, shaydulin_evaluating_2019}. In \figref{fig:concentration} of the main text, we showed the probability density of the QAOA parameter values after optimization instance by instance. Analytical arguments \cite{brandao_for_2018, streif_training_2019} suggest that the concentration of good parameter values may be affected by small size effects (here a small number of vertices of the graph to partition). A way to observe this effect is to quantify how the probability density changes when the number of vertices is increased from 12 to 16 and from 16 to 20. \figref{fig:concentration-diff} shows the absolute difference in the probability density for four types of graphs, from top to bottom: random 3-regular and 9-regular graphs, Erd\H{o}s graphs with $p_E=0.8$, and random weight graphs.
While the findings are hardly conclusive, we observe an overall decrease of the difference in the probability density, consistent with the expectation of a convergence in the limit $N\rightarrow \infty$.

\end{document}